\documentclass[twocolumn,showpacs,preprintnumbers,amsmath,amssymb]{revtex4}
\usepackage{mathrsfs}
\usepackage{amssymb}

\usepackage{graphicx}% Include figure files
\usepackage{bm}% bold math

\begin{document}

\title{Decoy State Quantum Key Distribution
With Modified Coherent State}% Force line breaks with \\

\author{Zhen-Qiang Yin, Zheng-Fu Han*, Fang-Wen Sun, Guang-Can Guo}
\affiliation{Key Lab of Quantum Information, CAS, USTC, China }

\date{\today}% It is always \today, today,

\begin{abstract}
To beat PNS attack, decoy state quantum key distribution (QKD) based
on coherent state has been studied widely. We present a decoy state
QKD protocol with modified coherent state (MCS). By destruction
quantum interference, MCS with fewer multi-photon events can be get,
which may improve key bit rate and security distance of QKD. Through
numerical simulation, we show about 2-dB increment on security
distance for BB84 protocol.

\end{abstract}
\maketitle

%\section{Introduction}
\section{Introduction}
Quantum Key Distribution (QKD) \cite{BB84,ekert1991,Gisin},
combining  quantum mechanics and conventional cryptography, allows
two distant peers (Alice and Bob) share secret string of bits,
called key. Any eavesdropping attempt to QKD process will introduce
high bit error rate of the key. By comparing part of the key, Alice
and Bob can catch any eavesdropping attempt. However, most of QKD
protocols, such as BB84, needs single photon source which is not
practical for present technology. Usually, real-file QKD set-ups
\cite{qkd1,qkd2,qkd3,qkd4,F-M} use attenuated laser pulses (weak
coherent states) instead. It means the laser source is equivalent to
a laser source that emits n-photon state $|n\rangle$ with
probability $P_n=\frac{\mu^n}{n!}e^{-\mu}$,where $\mu$ is average
photon number. This photon number Poisson distribution stems from
the coherent state $|\sqrt{\mu} e^{i\theta}\rangle$ of laser pulse.
Therefore, a few multi-photon events in the laser pulses emitted
from Alice open the door of Photon-Number-Splitting attack (PNS
attack) \cite{PNS1,PNS2,PNS3} which makes the whole QKD process
insecure. Fortunately, decoy state QKD theory \cite{decoy
theory1,decoy theory2,decoy theory3,decoy theory4,decoy theory5}, as
a good solution to beat PNS attack, has been proposed. And some
prototypes of decoy state QKD have been implemented \cite{decoy
experiment1,decoy experiment2,decoy experiment3,decoy
experiment4,decoy experiment5,decoy experiment6,decoy experiment7}.
The key point of decoy state QKD is to calculate the lower bound of
counting rate of single photon pulses ($S_1^L$) and upper bound of
quantum bit error rate (QBER) of bits generated by single photon
pulses ($e_1^U$). The tighter these bounds are given; longer
distance and higher key bit rate may be acquired. So a simple
question is how we can increase key bit rate and security distance
of decoy state QKD. Many methods to solve this question have been
presented, including more decoy states \cite{decoy theory5},
nonorthogonal decoy-state method \cite{nonorthogonal state
protocol}, photon-number-resolving method
\cite{photon-number-resolving method}, herald single photon source
method \cite{herald1,herald2}. Most of these methods are still based
on that photon number statistics obeyed Poisson distribution. From
derivation of formulas for estimating $S_1^L$ and $e_1^U$
\cite{decoy theory1, decoy theory2}, we know that the difference
between the real value of $S_1^L$ and $e_1^U$ origins from the
negligence of multi-photon counts events. Given some new laser
sources which have photon-number statistic distribution with less
probability of multi-photon events, a more precision estimation of
$S_1^L$ and $e_1^U$ should be obtained.

 In fact, it's proven that modified coherent state (MCS) with less probability of multi-photon
events could improve the security of QKD by \cite{MCS1}. The scheme
of MCS generation \cite{MCS2} relies on quantum interference to
depress multi-photon events from the coherent state. We can write
the MCS by \cite{MCS1} :

\begin{equation}
\begin{aligned}
|\Psi\rangle_{MCS}=\hat
{\mathcal{U}}|\alpha\rangle=\sum_{n=0}^{\infty}C_n|n\rangle
\end{aligned}
\end{equation}

with

\begin{equation}
\begin{aligned}
\hat {\cal U} =  \exp {1\over 2} (\zeta^*  \hat a^2  - \zeta  \hat
a^{\dagger 2})
\end{aligned}
\end{equation}

\begin{equation}
\begin{aligned}
C_n = {1\over \sqrt{n! \mu}} \Big ({\nu\over 2\mu}\Big)^{n\over
2}\exp\Big({\nu^*\over 2\mu}\alpha^2 - {|\alpha|^2\over 2}\Big)
H_n\Big({\alpha \over \sqrt{2\mu\nu}}\Big)
\end{aligned}
\end{equation}

\begin{equation}
\begin{aligned}
P_n=|C_n|^2
\end{aligned}
\end{equation}

and

\begin{eqnarray}
\mu \equiv \cosh(|\zeta|), ~\nu \equiv {\zeta\over|\zeta|}
\sinh(|\zeta|),~~ {\rm or}~ \mu^2= 1+|\nu|^2.\nonumber
\end{eqnarray}
with $\zeta$ is proportional to the amplitude of the pump field.

 In equation (3), $H_n$ represents the nth-order Hermite polynomial.
When $\alpha^2=\mu\nu$ ($\alpha^2=3\mu\nu$), the two-photon
(three-photon) events have been canceled. In followings, we always
assume $\alpha^2=c\mu\nu$ and $c$ is a positive constance. Like
conventional decoy state QKD based on coherent state, we rewrite the
density matrix of the source by introducing the randomization of
phase:

\begin{equation}
\begin{aligned}
|\rho_\nu\rangle&=\frac{1}{2\pi}\int_0^{2\pi}|\Psi\rangle_{MCS}\langle\Psi|
=\frac{1}{2\pi}\int_0^{2\pi}\hat{\mathcal{U}}||\alpha|e^{i\theta}\rangle\langle|\alpha|e^{i\theta}|d\theta\\
&=\sum_{n=0}^\infty P_n|n\rangle\langle n|
\end{aligned}
\end{equation}

Here, we can simply take $\alpha$, $\mu$, and $\nu$ as real number
because the value of $P_n$ only concerns with the module of them.
From equation (5), we can conclude that the MCS source is a source
that emits n-photon state $|n\rangle$ with probability $P_n$.

\section{Derivation}
 And now we can deduce formulas for 3-intensity MCS decoy QKD and 2-intensity MCS one.
Through adjusting the intensities of input coherent states
$|\alpha\rangle$, we can get sources of different $\nu$
corresponding to different $\alpha$. Two different sources of
density matrices $\rho_\nu$ and $\rho_{\nu'}$ could be get by this
way. The counting rates for the two sources ($\nu<\nu'$) are given
by:

\begin{equation}
\begin{aligned}
S_\nu=\sum_{n=0}^\infty P_n(\nu)S_n
\end{aligned}
\end{equation}

\begin{equation}
\begin{aligned}
S_{\nu'}=\sum_{n=0}^\infty P_n({\nu'})S_n
\end{aligned}
\end{equation}
where, $S_n$ represents counting rate for photon number state
$|n\rangle$. And quantum bit error rate (QBER) for $\nu'$ is:
\begin{equation}
\begin{aligned}
E_{\nu'}S_{\nu'}=\sum_{n=0}^\infty e_n P_n(\nu') S_n
\end{aligned}
\end{equation}
In which, $e_n$ is QBER for the key bits generated by photon number
state $|n\rangle$. To derive formulas for $S_1^L$ and $e_1^U$, it's
necessary to prove that $\frac{P_2(\nu')}{P_2(\nu)}Pn(\nu)\leqslant
P_n(\nu')$ for all of $n\geqslant 2$.

\begin{equation}
\begin{aligned}
&\frac{P_2(\nu')}{P_n{(\nu')}}-\frac{P_2(\nu)}{P_n{(\nu})}\\
&=\frac{2^{n-2}n!|H_2(\frac{1}{\sqrt{c}})|^2}{3!|H_n(\frac{1}{\sqrt{c}})|^2}((1+\frac{1}{\nu'^2})^{\frac{n-2}{2}}-(1+\frac{1}{\nu^2})^{\frac{n-2}{2}})
&\leqslant 0
\end{aligned}
\end{equation}
 From equation (9), we have proven $\frac{P_2(\nu')}{P_2(\nu)}P_n(\nu)\leqslant
 P_n(\nu')$. Now we can deduce the formulas for calculating $S_1^L$:

\begin{equation}
\begin{aligned}
S(\nu')&=P_0(\nu')S_0+P_1(\nu')S_1+P_2(\nu')S_2+P_3(\nu')S_3+\cdots\\
&\geqslant
P_0(\nu')S_0+P_1(\nu')S_1+\frac{P_2(\nu')}{P_2(\nu)}\sum_{n=2}^{\infty}P_n(\nu)S_n
\end{aligned}
\end{equation}
Combining with equation (6), we have

\begin{equation}
\begin{aligned}
S_1^L=\frac{(P_2(\nu)P_0(\nu')-P_2(\nu')P_0(\nu))S_0+P_2(\nu')S(\nu)-S(\nu')}{P_2(\nu')P_1(\nu)-P_2(\nu)P_1(\nu')}
\end{aligned}
\end{equation}
According to equation (8), estimation of $e_1^U$ is given by:

\begin{equation}
\begin{aligned}
e_1^U=\frac{(E_{\nu'}S_{\nu'}-\frac{S_0P_0(\nu')}{2})}{P_1(\nu')S_1^L}
\end{aligned}
\end{equation}

 Now we have get the formulas for calculating $S_1^L$ and $e_1^U$ for
three-intensity case. In this case Alice randomly emits laser pulses
from source $\rho_\nu$, $\rho_{\nu'}$, or doesn't emit anything,
then Bob can get counting rates for the three case: $S_\nu$,
$S_{\nu'}$ and $S_0$. Then Alice and Bob perform error correction
and private amplification by $S_1^L$ and $e_1^U$ calculated through
equation (11) and (12). The lower bound of security key rate is
given by \cite{decoy theory2}:

\begin{equation}
\begin{aligned}
R^L=q\{-S_{\nu'}f(E_{\nu'})H_2(E_{\nu'})+P_1(\nu')S_1^L[1-H_2(e_1^U)]\}
\end{aligned}
\end{equation}
with $q=\frac{1}{2}$ for BB84, $f(E_{\nu'})$ is he bidirectional
error correction efficiency (typically, $f(E_{\nu'})=1.2$), and
$H_2$ is the binary Shannon information function.

 For two-intensity case, Alice randomly emits laser pulses from source $\rho_\nu$ and
$\rho_{\nu'}$, then Bob can get counting rates for the two cases:
$S_\nu$ and $S_\nu'$. Now, Alice and Bob can get $S_0^U$ firstly,
then calculates $S_1^L$ by equation (14) with taking $S_0^U$ as
$S_0$. The formula for calculating $S_0^U$ can be derived from
equation (8) simply, it's:

\begin{equation}
\begin{aligned}
S_0^U=\frac{2E_{\nu'}S_{\nu'}}{P_0(\nu')}
\end{aligned}
\end{equation}
So the formula for two-intensity case is given by:

\begin{equation}
\begin{aligned}
&S_1^L\\
&=\frac{2(P_2(\nu)P_0(\nu')-P_2(\nu')P_0(\nu))E_{\nu'}S_{\nu'}+P_2(\nu')S(\nu)-S(\nu')}{(P_2(\nu')P_1(\nu)-P_2(\nu)P_1(\nu'))P_0(\nu')}
\end{aligned}
\end{equation}

 To get $e_1^U$, we can assume $S_0^L=0$ and let $S_0=S_0 ^L$,
then from equation (12) $e_1^U$ could be get by:

\begin{equation}
\begin{aligned}
e_1^U=\frac{E_{\nu'}S_{\nu'}}{P_1(\nu')S_1^L}
\end{aligned}
\end{equation}

 Equation (11) and (12) are formulas for three-intensity protocol, while equation
(15) and (16) are for two-intensity protocol. These are main results
of our derivation.

\section{improvement for decoy state QKD}
 In this section, our purpose is to show MCS's improvement for decoy state
QKD by numerical simulation. We consider the case when there is no
Eve. And from \cite{decoy theory4}:
$e_n=\frac{\frac{S_0}{2}+e_{det}\eta_n}{S_n}$,
$\eta_n=1-(1-\eta)^n$, $\eta=10^{-kL/10}\eta_{Bob}$,
$S_n=S_0+\eta_n$, where, $e_{det}$ is the probability that the
survived photon hits a wrong detector, $\eta$ is overall yield and
$\eta_n$ is yield for photon number state $|n\rangle$, $k$ is
transmission fiber loss constance, L is fiber length and
$\eta_{Bob}$ is the transmittance loss in Bob's security zone.
According to \cite{F-M}, we set $e_{det}=0.0135$, $S_0=8\times
10^{-7}$, $k=0.2dB/Km$ for numerical simulation. We simply set
$\eta_{Bob}=1$, because our purpose is a comparison not absolute
distance. These are our parameters and formulas for numerical
simulation.
\subsection{To Cancel Two-photon Events}
 Here, we set $c=1$ to cancel all two-photon events. We cannot use
equation (11) and (12) immediately for $P_2=0$. But, it's easy to
see that we can replace $P_2$ as $P_3$, and now the equations are
available for this case.
 Firstly, we will show the increment of precision for estimating $S_1^L$.
Typically, we set $\alpha=\sqrt{0.2}$, $\alpha'=\sqrt{0.6}$ as the
two inputs for the MCS generator. With these inputs, one can get two
kinds of MCS with $\nu=0.196$ and $\nu'=0.53$. Fig1 shows that real
$S_1$, $S_1^L$ calculated by ordinary decoy state QKD based on
coherent state ($\alpha=\sqrt{0.2}$ for decoy state and
$\alpha'=\sqrt{0.6}$ for signal state) and $S_1^L$ calculated by MCS
decoy QKD ($\nu$=0.196 for the decoy state and $\nu'$=0.53 for the
signal state). From Fig1, we can conclude that for two-intensity
case MCS decoy state QKD is indeed more effective to calculate
$S_1^L$ than traditional decoy state QKD based on coherent state. We
found that in two-intensity protocol the longest length still
capable of estimating $S_1^L$ precision increases by 20KM.

\begin{figure}[!t]\center
\resizebox{7.5cm}{!}{\includegraphics{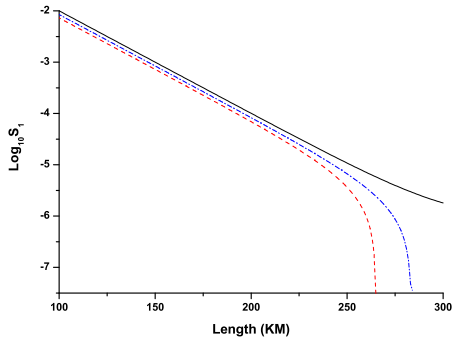}} \caption{Couting
rate for single-photon laser pulses ($S_1$) verse fiber length $L$.
Solid curve: real value of counting rate for single-photon laser
pulses for no eavesdropping case. Dashed curve: $S_1^L$ calculated
by traditional decoy state QKD based on coherent state.
Dotted-dashed curve: $S_1^L$ calculated by MCS decoy
QKD.}\label{schematic}
\end{figure}

 Secondly, we compare the key bit rate $R$ of MCS decoy QKD and
QKD based on coherent state. To compare the two decoy QKD process
more fairly, we draw Fig2 in which each point has optimal value of
$\alpha$ and $\alpha'$ or $\nu$ and $\nu'$ for two-intensity case.
But for three-intensity case, we set the average photon-number of
decoy pulses as 0.1 and $\nu'$ or $\alpha'$ has optimal value for
each point.

\begin{figure}[!h]\center
\resizebox{7.5cm}{!}{\includegraphics{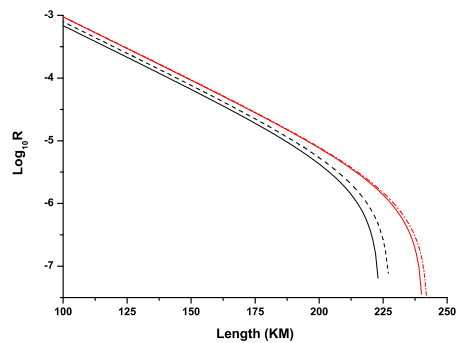}}
\caption{security key rate ($R^L$) verse fiber length $L$. Solid
curves: $R^L$ with 2-intensity and 3-intensity decoy QKD based on
coherent state. Dashed curves: $R^L$ with 2-intensity and
3-intensity MCS decoy QKD.} \label{schematic}
\end{figure}

 From Fig2, we see that in two-intensity case about 3KM
increment on security distance could be get by using MCS and in
three-intensity case 2KM increment is given.

\subsection{To Cancel Three-photon Events}
 Here, we set $c=3$ to cancel all two-photon events. And equation
(11) and (12) can be used immediately. Though MCS without
three-photon events has more multi-photon events than the one
without two-photon events, former has higher total counting rates
and lower QBER, which may increase $R$. The results are drawn in
Fig3. In Fig3, each point has optimal value of $\alpha$ and
$\alpha'$ or $\nu$ and $\nu'$ for two-intensity case. But for
three-intensity case, we set the average photon-number of decoy
pulses is 0.1 and $\nu'$ or $alpha'$ has optimal value for each
point.

\begin{figure}[!h]\center
\resizebox{7.5cm}{!}{\includegraphics{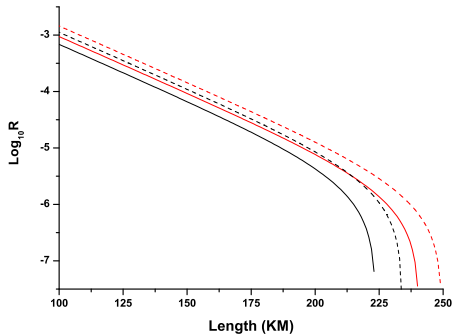}}
\caption{security key rate ($R^L$) verse fiber length $L$. Solid
curves: $R^L$ with 2-intensity and 3-intensity decoy QKD based on
coherent state. Dashed curves: $R^L$ with 2-intensity and
3-intensity MCS decoy QKD.}\label{schematic}
\end{figure}

 From Fig3, we see nearly 2-dB increment is given on security
length both for two states protocol and three states protocol. This
result is better than $c=1$ MCS. We found MCS without three-photon
events has higher counting rates and lower QBER than MCS (c=1). This
is the reason why $c=3$ MCS has better performance.

 In above discussion, we set $c=1$ to cancel two-photon events or
$c=3$ to cancel three-photon events. However, we can also set $c$ as
some arbitrary positive value, provided this value make $R$ rise.
And we draw Fig4 in which the relation of increment of security
distance between $c$ is given. From Fig4, we see optimal $c$ is
different for two-intensity and three intensity cases. For
two-intensity case, the optimal value is 3.3 and for three-intensity
it's 2.8.

\section{Conclusion}

 According to above discussion, we see that: thanked to MCS's fewer
multi-photon events probability, decoy state with MCS source can
indeed provide QKD service of higher key bit rate and longer
distance than before. We found about 2-dB increment of security
distance is acquired. Generating this kind of MCS laser pulses isn't
difficult for today's Lab. We expect that our MCS decoy QKD scheme
could be implemented earlier.

 The authors thank Prof.Qing-Yu Cai for his helpful advice. This work
was supported by National Fundamental Research Program of China
(2006CB921900), National Natural Science Foundation of China
(60537020,60621064) and the Innovation Funds of Chinese Academy of
Sciences. To whom correspondence should be addressed, Email:
zfhan@ustc.edu.cn.

\begin{figure}[!h]\center
\resizebox{7.5cm}{!}{\includegraphics{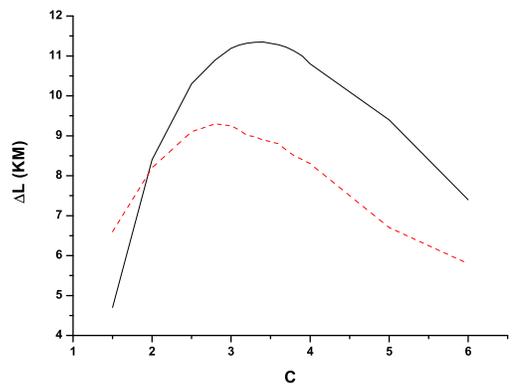}} \caption{increment of
security distance ($\Delta L$) verse $c$. Solid curve: for the
2-intensity case. Dotted curve: for the 3-intensity case. And each
point has optimal $\nu$ and $\nu'$.}\label{schematic}
\end{figure}


\begin{thebibliography}{00}


\bibitem{BB84}
C. H. Bennett,  G.Brassard, Proceedings of \emph{IEEE International
Conference on Computers, Systems, and Signal Processing}, (IEEE,
1984), pp. 175-179.
\bibitem{ekert1991}
A. K. Ekert, \emph{Phys. Rev. Lett.} \textbf{67} 661 (1991)
\bibitem{Gisin}
N. Gisin, Gr¨¦goire Ribordy, W. Tittel, and H. Zbinden Rev.
Mod.Phys. \textbf{74}, 145 (2002)
\bibitem{qkd1}
M. Bourennane et al., Opt. Express \textbf{4}, 383 (1999)
\bibitem{qkd2}
D. Stucki et al., New. J. Physics, \textbf{4}, 41, (2002)
\bibitem{qkd3}
H. Kosaka et al., Electron. Lett. \textbf{39}, 1199 (2003)
\bibitem{qkd4}
C. Gobby, Z.L. Yuan, and A.J. Shields, Appl. Phys. Lett.
\textbf{84}, 3762 (2004);
\bibitem{F-M}
X.-F. Mo, B. Zhu, Z.-F. Han, Y.-Z. Gui, G.-C. Guo Optics Letters,
Vol. \textbf{30}, Issue 19, pp. 2632-2634 (October 2005)
\bibitem{PNS1}
B. Huttner, N. Imoto, N. Gisin, and T. Mor, Phys. Rev. A
\textbf{51}, 1863 (1995);
\bibitem{PNS2}
G. Brassard, N. Lu¡§tkenhaus, T. Mor, and B. C. Sanders, Phys. Rev.
Lett. \textbf{85}, 1330 (2000).
\bibitem{PNS3}
N. Lu¡§tkenhaus, Phys. Rev. A \textbf{61}, 052304 (2000).
\bibitem{decoy theory1}
W.-Y. Hwang, Phys. Rev. Lett. \textbf{91}, 057901 (2003).
\bibitem{decoy theory2}
H.-K. Lo, X. Ma, and K. Chen, Phys. Rev. Lett. \textbf{94}, 230504
(2005).
\bibitem{decoy theory3}
X.-B. Wang, Phys. Rev. Lett. \textbf{94}, 230503 (2005);
\bibitem{decoy theory4}
X. Ma et al., Phys. Rev. A \textbf{72}, 012326 (2005).
\bibitem{decoy experiment1}
Y. Zhao, B. Qi, X. Ma, H.-K. Lo, and L. Qian Phys. Rev. Lett.
\textbf{96}, 070502 (2006)
\bibitem{decoy experiment2}
Yi Zhao, Bing Qi, X.-F. Ma, H.-K. Lo, Li Qian Proceedings of IEEE
International Symposium on Information Theory 2006, pp. 2094-2098
\bibitem{decoy experiment3}
C.-Z. Peng, J. Zhang, D. Yang, W.-B. Gao, H.-X. Ma, H. Yin, H.-P.
Zeng, T. Yang, X.-B. Wang, and J.-W. Pan Phys. Rev. Lett.
\textbf{98}, 010505 (2007)
\bibitem{decoy experiment4}
D. Rosenberg, J. W. Harrington, P. R. Rice, et al., Phys. Rev. Lett.
\textbf{98}, 010503 (2007)
\bibitem{decoy experiment5}
Z. L. Yuan, A. W. Sharpe, and A. J. Shields Appl. Phys. Lett.
\textbf{90} 011118 (2007).
\bibitem{decoy experiment6}
Tobias Schmitt-Manderbach et al. Phys. Rev. Lett. \textbf{98},
010504 (2007)
\bibitem{decoy experiment7}
Z.-Q. Yin, Z.-F. Han, Wei Chen, F.-X. Xu, Q.-L. Wu, G.-C. Guo
quant-ph/0704.2941 (2007)
\bibitem{decoy theory5}
X.-B. Wang, Phys. Rev. A \textbf{72}, 012322 (2005)
\bibitem{nonorthogonal state protocol}
J.-B. Li, and X.-M. Fang, arXiv:quant-ph/0509077 (2005)
\bibitem{photon-number-resolving method}
Qing-yu Cai, and Yong-gang Tan Phys. Rev. A. \textbf{73}, 032305
(2006)
\bibitem{herald1}
Tomoyuki Horikiri and Takayoshi Kobayashi Phys. Rev. A \textbf{73},
032331 (2006)
\bibitem{herald2}
Qin Wang, X.-B. Wang, and G.-C. Guo Phys. Rev. A \textbf{75}, 012312
(2007)
\bibitem{MCS1}
Y.-J. Lu and Z.-Y. Ou, Phys. Rev. A \textbf{71}, 032315 (2005)
\bibitem{MCS2}
Y.-J. Lu and Z.-Y. Ou, Phys. Rev. Lett \textbf{88}, 023601 (2002)
\end{thebibliography}
\end{document}